\documentstyle[11pt]{article}
\newcommand{\beq}{\begin{equation}}
\newcommand{\eeq}{\end{equation}} 

\newcommand{\beqa}{\begin{eqnarray}}
\newcommand{\eeqa}{\end{eqnarray}}

\def\half{\frac{1}{2}}
\def\opone{\leavevmode\hbox{\small1\kern-3.8pt\normalsize1}}

\begin{document}

\title{Bell inequality for arbitrary many settings of the analyzers}
\author
{N. Gisin\\
\protect\small\em Group of Applied Physics, University of Geneva, 
1211 Geneva 4, Switzerland}
\date{\today}

\maketitle

\begin{abstract}
A generalization of the CHSH-Bell inequality to arbitrary many settings is presented.
The singlet state of two spin $\half$ violates this inequality for all numbers
of setting. In the limit of arbitrarily large number of settings, 
the violation tends to the finite ratio $\frac{4}{\pi}\approx 1.27$.
\end{abstract}

\section{Introduction}\label{introduction}
One of the most remarkable feature of our basic physical science, quantum mechanics,
is certainly its holistic description of Nature. In particular, when we divide her into parts,
then, according to quantum theory, these parts are generally entangled. As a consequence,
what happens here can't be considered as independent of what happens there, as demonstrated
by Bell's theorem (i.e. that entangled quantum states violate Bell inequality) \cite{Bell64}. 
Considering
the importance of quantum non-locality, both for the worldview quantum physics presents and for
the potential applications of entanglement in quantum computing and communication \cite{QIPintro98}, it is
important to examine the basis of entanglement as closely as possible. Let us, for example,
notice that despite that Bell's result is now 35 years old, no loophole free experiment
has yet been realized \cite{detloophole,NGBG99}!

The most well-known Bell inequality is the CHSH inequality \cite{CHSH69}:
\beq
S2=E(\vec a,\vec b)+E(\vec a,\vec b')+E(\vec a',\vec b)-E(\vec a',\vec b') \le 2
\label{CHSH}
\eeq
This inequality refers to 2 2-dimensional systems (spin $\half$) with 2 settings on each of the 2 sides.
Clearly, there are many possible generalizations. Generalizations to n 2-dimensional systems with 2 settings are known
\cite{Bellnqubit98}.
In this short letter, we present a generalization to 2 2-dimensional systems with n settings.

\section{CHSH-Bell inequality for n settings}
Let $a_j=\pm1$ and $b_j=\pm1$ for all indices $j=1...n$. The following inequalities can be
easily checked by inspection:
\beqa
\label{ineq2}
&a_1&(b_1+b_2) \\ \nonumber
+&a_2&(b_1-b_2) \le 2
\eeqa
\beqa
\label{ineq3}
&a_1&(b_1+b_2+b_3) \\ \nonumber
+&a_2&(b_1+b_2-b_3) \\ \nonumber
+&a_3&(b_1-b_2-b_3) \le 5
\eeqa
\beqa
\label{ineq4}
&a_1&(b_1+b_2+b_3+b_4) \\ \nonumber
+&a_2&(b_1+b_2+b_3-b_4) \\ \nonumber
+&a_3&(b_1+b_2-b_3-b_4) \\ \nonumber
+&a_4&(b_1-b_2-b_3-b_4) \le 8
\eeqa
Note that minus signs appear only below the diagonal.

From the first of these inequalities, i.e. from inequality (\ref{ineq2}), the usual CHSH-Bell
inequality (\ref{CHSH}) can be deduced with 
$E(\vec a,\vec b) =\int \rho(\lambda)d\lambda~ a(\vec a,\lambda)~b(\vec b,\lambda)$,
where $a(\vec a,\lambda)=\pm1$ is the outcome given the setting $\vec a$ and the local
hidden variable $\lambda$ and similarly for $b(\vec b,\lambda)=\pm1$ 
(note the important assumption that $a(\vec a,\lambda)$ and $b(\vec b,\lambda)$ are independent
of the other setting $\vec b$ and $\vec a$, respectively). 

From the inequalities (\ref{ineq3}) and (\ref{ineq4}), the following generalization of the Bell-CHSH inequality obtains,

\beqa
S3\equiv&E(\vec a_1,\vec b_1)&+E(\vec a_1,\vec b_2)+E(\vec a_1,\vec b_3)  \\ \nonumber
       +&E(\vec a_2,\vec b_1)&+E(\vec a_2,\vec b_2)-E(\vec a_2,\vec b_3)  \\ \nonumber
       +&E(\vec a_3,\vec b_1)&-E(\vec a_3,\vec b_2)-E(\vec a_3,\vec b_3) \leq 5
\eeqa
\beqa
S4\equiv&E(\vec a_1,\vec b_1)&+E(\vec a_1,\vec b_2)+E(\vec a_1,\vec b_3)+E(\vec a_1,\vec b_4)  \\ \nonumber
       +&E(\vec a_2,\vec b_1)&+E(\vec a_2,\vec b_2)+E(\vec a_2,\vec b_3)-E(\vec a_2,\vec b_4)  \\ \nonumber
       +&E(\vec a_3,\vec b_1)&+E(\vec a_3,\vec b_2)-E(\vec a_3,\vec b_3)-E(\vec a_3,\vec b_4)  \\ \nonumber
       +&E(\vec a_4,\vec b_1)&-E(\vec a_4,\vec b_2)-E(\vec a_4,\vec b_3)-E(\vec a_4,\vec b_4) \leq 8
\eeqa
This set of inequalities generalizes to arbitrary many settings of the analyzers:
\beq
Sn\equiv\sum_{j=1}^n\left(\sum_{k=1}^{n+1-j} E(\vec a_j,\vec b_k)-\sum_{k=n+2-j}^n E(\vec a_j,\vec b_k)\right)
\leq[\frac{n^2+1}{2}]\equiv Sn_{lhv}
\label{ineqn}
\eeq
where $[x]$ denotes the largest integer smaller or equal to $x$. The above inequality
(\ref{ineqn}) is the main result of this letter.

For any product state and any number of settings $n$, there are settings that
saturate inequality (\ref{ineqn}) \cite{nonmaxent}. 
On the opposite, for maximal entanglement, like in the singlet state $\psi$, the
inequality (\ref{ineqn}) can be violated by quantum correlations. Numerical
evidence shows that it suffice to consider co-planar settings $\vec a_j$ and $\vec b_j$:
$E(\alpha,\beta)=<\vec a_j\vec\sigma\otimes\vec b_k\vec\sigma>_{\psi}=-\cos(\alpha-\beta)$, 
where $\vec a_j=(\sin(\alpha_j),0,\cos(\alpha_j))$ and $\vec b_j=(\sin(\beta_j),0,\cos(\beta_j))$.
Indeed, the following settings, $j=1,...,n$:
\beq
\alpha_j=j\frac{\pi}{n}  \hspace{1 cm}  \beta_j=\frac{3-n-2j}{2}\frac{\pi}{n}
\eeq
lead to the maximal violation of (\ref{ineqn}):
\beq
Sn_{QM}=2n\frac{\cos(\frac{\pi}{2n})}{\sin(\frac{\pi}{n})} > [\frac{n^2+1}{2}] = Sn_{lhv}
\eeq
The ratio of violation is shown in Fig. 1 as a function of the number of settings.
Asymptotically, i.e. for large n, the sums  in (\ref{ineqn}) 
converge into integrals and the ratio of violation tends to:
\beq
\frac{S\infty_{QM}}{S\infty_{lhv}}=
\lim_{n\rightarrow\infty}\frac{2n\frac{\cos(\frac{\pi}{2n})}{\sin(\frac{\pi}{n})}}{\frac{n^2+1}{2}}
=\frac{4}{\pi}\approx 1.273 > 1 
\eeq

\section{Conclusion}
A generalization of the CHSH-Bell inequality to arbitrary many settings has been presented.
For large numbers of settings, the ratio by which this generalized inequality is violated by
quantum mechanics is smaller than the usual $\sqrt{2}$ factor (i.e. 41\%) 
valid for 2 settings. However,
this ratio is still significant: $27.3\%$. Consequently, experimental data reproducing the
entire quantum correlation function could violate the generalized inequality by a larger
number of standard deviations, thanks to improved statistics \cite{Zukowski93}.

Another possible advantage of this generalized inequality could have been that it is
less sensitive to the detection loophole \cite{detloophole}. 
Admittedly, this was our original motivation for this work. However, it turns
out that the corresponding generalization for the CH inequality \cite{CH74} does not allow to lower the detector's
efficiency threshold necessary to close the detection loophole. Actually, recently we found a local hidden variable
model based on the detection loophole which reproduces exactly the quantum correlation function \cite{NGBG99}!
Consequently, this model does also violate the generalized inequality (\ref{ineqn}) for any number of settings.

\small
\section*{Acknowledgments}
This work profited from stimulating discussions with Bernard Gisin and Bruno Huttner. It 
was partially supported by the Swiss National Science Foundation and by
the European TMR Network "The Physics of Quantum Information" through the Swiss
OFES.

\section*{Figure Captions}
\begin{enumerate}
\item Ratio of the maximal violation of the inequality with increasing numbers of settings.
The usual CHSH-Bell inequality corresponds to 2 settings (violation ratio$=\sqrt{2}\approx1.41$). Assymptotically
the ratio tends to $\frac{4}{\pi}\approx1.273$.
\end{enumerate}


\begin{thebibliography}{99}
\bibitem{Bell64} J. S. Bell, Physics {\bf1}, 195 (1964).
\bibitem{QIPintro98} {\it Introduction to Q computation and information}, eds H.K. Lo,
       S. Popescu and T. Spiller, World Scientific 1998.
\bibitem{detloophole} P. Pearle, Phys. Rev. D {\bf2}, 1418 (1970); 
       J.F. Clauser, M.A. Horne, A. Shimony, and R.A. Holt, Phys. Rev. Lett., 
                 {\bf23}, 880, (1969);
        E. Santos, Phys.Rev. A, {\bf46}, 3646, (1992). The following proposed experiments aim
        at closing the detection loophole: P. Kwiat et al., Phys. Rev. A {\bf49}, 3209, (1994);
        E.S. Fry, T. Walther and S. Li, Phys. Rev. A {\bf52}, 4381, (1995);
        Huelga et al., Phys. Rev. A {\bf51}, 5008, (1995);
\bibitem{NGBG99} N. Gisin and B. Gisin, quant-ph 9905018 (see also the software demo at 
       http://www.gap-optique.unige.ch/News/BellSoft.asp).
\bibitem{CHSH69}J. F. Clauser, M. A. Horne, A. Shimony, and R. A. Holt, Phys.
                Rev. Lett. {\bf23}, 880 (1969).
\bibitem{Bellnqubit98} N. Gisin and Helle Bechmann-Pasquinucci, Phys. Lett. A {\bf246}, 1, 1998; 
       and references there in.
\bibitem{nonmaxent} It is not clear whether arbitrary entangled states violate all the 
       inequalities (\ref{ineqn}) for some settings (as is the case for the CHSH inequality,
       see N. Gisin, Phys. Lett. A {\bf154}, 201, 1991). 
       The case of odd numbers of setting appears special. 

\bibitem{Zukowski93} M. \.Zukowsky, Phys. Lett. A {\bf177}, 290, 1993.
\bibitem{CH74}J. F. Clauser, and M. A. Horne, Phys. Rev. D {\bf10}, 526 (1974).




\end{thebibliography}
\end{document}